\documentclass[prd,nofootinbib,preprint,superscriptaddress]{revtex4}

\usepackage{amsmath, amssymb, amsthm, graphicx, epsfig, fancyhdr,epsfig, slashed}

\usepackage{tikzsymbols}
\usepackage{natbib}
\usepackage{float}

\usepackage{tikz,xcolor,hyperref}
\usepackage{slashed}
\usepackage{booktabs,array}

\definecolor{lime}{HTML}{A6CE39}
\DeclareRobustCommand{\orcidicon}{
	\begin{tikzpicture}
	\draw[lime, fill=lime] (0,0) 
	circle [radius=0.2] 
	node[white] {{\fontfamily{qag}\selectfont \tiny ID}};
	\draw[white, fill=white] (-0.0625,0.095) 
	circle [radius=0.007];
	\end{tikzpicture}
	\hspace{-2mm}
}

\foreach \x in {A, ..., Z}{\expandafter\xdef\csname orcid\x\endcsname{\noexpand\href{https://orcid.org/\csname orcidauthor\x\endcsname}
			{\noexpand\orcidicon}}
}

\newcommand{\be}{\begin{equation}}
\newcommand{\ee}{\end{equation}}
\newcommand{\bea}{\begin{eqnarray}}
\newcommand{\eea}{\end{eqnarray}}

\newcommand{\ba}{\begin{eqnarray}}
\newcommand{\ea}{\end{eqnarray}}
\newcommand{\bi}{\begin{itemize}}
\newcommand{\ei}{\end{itemize}}

\newcommand{\x}{\star}

\newcommand{\gw}[1]{\textcolor{orange}{[Graham: #1]}}

\begin{document}


\title{Complementarity between Cosmic String Gravitational Waves and long-lived particle searches in a laboratory }

\author{Satyabrata Datta\orcidA{}}
\email{amisatyabrata703@gmail.com}
\affiliation{Institute of Theoretical Physics and Institute of Physics Frontiers and Interdisciplinary Sciences, Nanjing Normal University}
\affiliation{Nanjing Key Laboratory of Particle Physics and Astrophysics, Nanjing 210023, China}
\author{Ambar Ghosal\orcidB{}}
\email{ambar.ghosal@gmail.com}
\affiliation{Saha Institute of Nuclear Physics, 1/AF, Bidhannagar, Kolkata 700064, India}

\author{Anish Ghoshal\orcidC{}}
\email{anish.ghoshal@fuw.edu.pl}
\affiliation{Institute of Theoretical Physics, Faculty of Physics, \\ University of Warsaw, ul. Pasteura 5, 02-093 Warsaw, Poland}

\author{Graham White\orcidD{}}
\email{graham.white@gmail.com}
\affiliation{School of Physics and Astronomy, University of Southampton, Southampton SO17 1BJ, United Kingdom}

\newcolumntype{C}[1]{>{\centering}m{#1}}

\begin{abstract}
\textit{
Cosmic strings are powerful witnesses to cosmic events including any period of early matter domination. If such a period of matter domination was catalysed by metastable, long-lived particles, then there will be complementary signals to ascertain the nature of dark sector in experiments detecting primordial features in the gravitational wave (GW) power spectrum and laboratory searches for long-lived particles. We give explicit examples of global and local U(1) gauge extended dark sectors to demonstrate such a complementarity as the union of the two experiments reveals more information about the dark sector than either experiment. Demanding that
Higgs-portal long-lived scalar be looked for, in various experiments such as DUNE, FASER, FASER-II, MATHUSLA,
and SHiP, we identify the parameter space that leads to complementary observables for GW detectors such as Laser Interferometer Space Antenna and Einstein Telescope.
}

\end{abstract}

\maketitle

\section{Introduction}


Networks of cosmic strings (CS) are ubiquitous in cosmological models that include grand unification, as many symmetry-breaking paths contain steps where the vacuum manifold has noncontractible loops \cite{Dror:2019syi,Dunsky:2021tih,King:2021gmj,Fu:2023mdu}. Such a network of CS continuously emits gravitational waves (GW), resulting in a spectrum that is flat during radiation domination, apart from the evolution of $g_\ast$ \cite{Cui:2018rwi}. Any departure from such a flat spectrum speaks to a surprise in our cosmic history \cite{Cui:2018rwi,Ghoshal:2023sfa,Blasi:2020wpy,Ferrer:2023uwz} with a period of early matter domination providing a particularly striking feature in the otherwise flat spectrum \cite{Cui:2017ufi,Ghoshal:2023sfa, Datta:2020bht, Samanta:2021zzk, Chianese:2024gee, Datta:2024bqp}. Such a feature reveals two observables which yield information on the culprit for the early matter domination - the time matter domination began, which is set by the highest frequency that departs from a flat spectrum, and the duration of matter domination, which corresponds to the width of the feature in frequency space.
A period of early matter domination can be caused by a metastable, long-lived particle \cite{Coughlan:1983ci,Starobinsky:1994bd,Dine:1995uk,Moroi:1999zb,Ghoshal:2022ruy,Bishara:2024rtp}. Such a particle can arise in dark sectors as a mediator \cite{Ghoshal:2022ruy}. The period of early matter domination arises when the metastable particle thermally decouples with an abundance much larger than the observed dark matter abundance. Since matter dilutes slower than radiation, as long as the particle is sufficiently long-lived there will be a period of matter domination. 

Such long-lived particles are notoriously difficult to detect in traditional colliders due to their displaced vertex making them invisible  \cite{Feng:2022inv}. If the particle is sufficiently light, $\lesssim 0.3$ GeV, there exist severe constraints from supernova cooling.\footnote{Recent studies indicate even more stringent bounds on the Higgs
mixed scalars decaying into ($e^+e^-$) pairs \cite{Caputo:2022mah,Caputo:2021rux}. Similarly,
 neutron star mergers like GW170817 impose constraints via
 charged lepton decays, generating detectable fireballs \cite{Diamond:2021ekg,Diamond:2023cto}. These constraints extend to $\sim$GeV masses and probe smaller
 mixing angles, considering energy trapping effects \cite{Fiorillo:2025yzf}.} For heavier particles, beam dump experiments and long-lived searches such as FASER, FASER-II \cite{FASER:2018bac,FASER:2019aik}, DUNE \cite{DUNE:2015lol} and MATHUSLA \cite{Curtin:2018mvb, Curtin:2023skh} promise to probe large parts of the parameter space. However, any signal would leave the hidden sector woefully under-determined with only the lifetime and mass of the particle known. It is therefore crucial to find complementary sources of knowledge about such a long-lived particle. From primordial features in the stochastic gravitational wave background, we will show that one may obtain an independent measurement of the lifetime as well as a measurement of the annihilation cross-section between hidden and visible sectors. For the case of a hidden scalar singlet with three free parameters, one can test the BSM model due to the synergy between gravitational wave detectors and laboratory searches, for instance as proposed in Ref.\cite{Ghoshal:2022ruy} in the context of inflationary GW. For generic studies on the impact of long-lived particles in inflationary tensor perturbations, see Ref.\cite{Berbig:2023yyy,Borboruah:2024eha,Borboruah:2024eal,Ghoshal:2024gai,Bernal:2020ywq, Datta:2022tab, Datta:2023vbs, Chianese:2024nyw}.  

In this work, we consider the thermal history of a long-lived particle where CS are in the background bearing witness to periods of early matter domination in order to demonstrate the possibility of complementarity. In section \ref{sec:model}, we give an explicit model. We then discuss the thermal history of our model in section \ref{sec:history}. Next, we give the imprint of CS onto the parameter space in section \ref{sec:strings} before discussing the ground-based signals in section \ref{sec:longlived}. Finally, we explicitly show the complementarity in section \ref{sec:comp} before concluding.

\section{The model}\label{sec:model}

Our main objective is to present a concrete scenario where complementary information about a model can be gleaned from ground-based experiments and the spectral shape of the GW background from CS, due to the presence of a particle sufficiently long-lived to catalyze a period of early matter domination. In order to realise such a non-standard cosmological evolution we consider the case where the Standard Model (SM) is extended by additional scalar fields. We also require a cosmic GW background generated by CS. Since a visible background requires a very high symmetry-breaking scale, the physics that produces such strings we assume to be weakly coupled from the additional scalars we will be discussing in detail. 
Keeping these objectives in view, the SM is extended by a $U(1)_{B-L}$ symmetry (with $B-L$ complex scalar $\Phi$) along with additional $\mathcal{Z}_2$-odd real scalar singlet, $S$. One can write the symmetry invariant full scalar potential as
\begin{equation}
\begin{aligned}
{} &    
V(H,S,\Phi)\supset\mu^2(H^\dagger H)-\lambda (H^\dagger H)^2+\mu_\Phi^2(\Phi^\dagger \Phi)-\lambda_\Phi(\Phi^\dagger \Phi)^2+\frac{1}{2}\mu_S^2 S^2-\frac{1}{4}\lambda_S S^4\\ &
-\frac{1}{2}\lambda_{\Phi S}(\Phi^\dagger \Phi)S^2-\frac{1}{2}\lambda_{HS}(H^\dagger H)S^2-\lambda_{\Phi H}(\Phi^\dagger \Phi)(H^\dagger H) \ .
\end{aligned}
\end{equation}
After breaking the electroweak (EW) symmetry, $U (1)_{B-L}$ symmetry, and our global discrete $\mathcal{Z}_2$ symmetry, the scalar potential becomes a function of three classical fields $H=\frac{1}{\sqrt{2}}\begin{bmatrix}
0 \\
v+h
\end{bmatrix}
$, $S=(v_s+s)$, and $\Phi=\frac{1}{\sqrt{2}}(v_{BL}+\phi)$, $v$ being the EW scale, such that
\begin{equation}
\begin{aligned}
{} &    
V(h,s,\phi)\supset\frac{1}{2}\mu^2(v+h)^2-\frac{1}{4}\lambda (v+h)^4+\frac{1}{2}\mu_\Phi^2(v_{BL}+\phi)^2-\frac{1}{4}\lambda_\Phi(v_{BL}+\phi)^4+\frac{1}{2}\mu_S^2 (v_s+s)^2\\ &
-\frac{1}{4}\lambda_S (v_s+s)^4-\frac{1}{4}\lambda_{\Phi S}(v_{BL}+\phi)^2(v_s+s)^2-\frac{1}{4}\lambda_{HS}(v+h)^2(v_s+s)^2-\frac{1}{4}\lambda_{\Phi H}(v_{BL}+\phi)^2(v+h)^2 \ .
\end{aligned}
\end{equation}
For the sake of analytic clarity, in this section, we will assume a hierarchy between the VEVs,
\begin{equation}
    v_{BL}\gg v_s \gg v , \quad m_H >> m_S \ .
\end{equation}
The choice of this hierarchy among the VEV scales and masses imposes certain possible constraints on the potential parameters: specifically, we have $\lambda \gg \lambda_s \gg \lambda_{\Phi S}$, and the feeble mixing angle translates to $\lambda_{HS}\ll \lambda$. One motivation for such hierarchies $v_{BL}\gg v_s \gg v$ and $m_S\ll m_H$, can be thought to be related to neutrino mass generation mechanisms and dark matter physics, where $v_{BL}$ controls the seesaw scale \cite{Mohapatra:1980qe}, which is typically very high for the vanilla type-I seesaw, where right-handed neutrinos have masses of the $\mathcal{O}(10^{12}$ GeV) or so. The scale $v_s$ is associated with dark sector, particularly in the context of freeze-in dark matter. Such a scalar whose search we propose in the complementary manner as long-lived particle then becomes a Higgs-portal like mediator for the dark matter, already discussed, see Refs. \cite{Ghoshal:2022ruy,Barman:2021lot}.
This of course is one corner of parameter space that features in our scan, where we make no such restriction. 
Interestingly, the feeble effective coupling between the $B-L$ scalar and such a real scalar $S$ of the form
\begin{equation}
V_{\rm int}(\Phi, S)\supset -\frac{1}{2}\lambda_{\Phi S} (\Phi^\dagger \Phi) S^2 +\frac{\lambda_S}{4} S^4
\end{equation}
can generate such a $B-L$ Higgs dependent VEV, $v_s=\sqrt{\lambda_{\Phi S}/\lambda_S} v_{BL}$ 
and following the EW symmetry breaking one can treat the model as a 2-field system having a mixed portal, controlled by the coupling $\lambda_{HS}$ thus providing an effective three-body interaction, as well as a mixing term \begin{equation}
    V(H,S)\supset\frac{1}{2} \lambda_{HS}vHS^2 + \frac{1}{2}\lambda_{HS} v v_s H S ,
\end{equation}
such that the singlet can only decay via mixing with the Higgs.
The mixing and the decay rate are controlled by a mixing angle which we derive by diagonalizing the $2 \times 2$ mass matrix
\begin{equation}
\mathcal{M}^2=
\begin{bmatrix}
2\lambda v^2 & \lambda_{HS} v v_s \\
\lambda_{HS} v v_s  & 2 \lambda_S v_s^2 \ 
\end{bmatrix} .  
\end{equation}
After diagonalization, one finds the mass eigenvalues,
\begin{equation}
\begin{aligned}
{} &  
m_{H}\simeq\sqrt{2\lambda} v \sim 125 \text{\:GeV}\\ &
m_{S}\simeq\sqrt{2\lambda_S} v_s\\ &
\tan\theta\sim \frac{\lambda_{HS} v v_s}{m_H^2-m_S^2}.
\end{aligned}
\end{equation}
The singlet mass and the mixing angle, $\theta$ will be the two variables that we will reconstruct from ground-based and GW observables.
\medskip

\section{Cosmic history}\label{sec:history}

In this section, we introduce the ingredients required to understand the cosmic history with an early matter domination (EMD). The additional scalar, $S$, can decouple from the thermal bath early on, eventually dominating the energy density as a non-relativistic matter component. This can cause a period of early matter domination before \(S\) eventually decays primarily into the visible sector. The interaction rate of \(S\) with the visible sector is governed by the thermally averaged annihilation cross-section multiplied by the relative velocity, \(\langle \sigma_S v \rangle\). The evolution of the number density of \(S\) is described by the Boltzmann equation:
\begin{equation}
\frac{d n_S}{dt} + 3{H} n_S = \langle \sigma_S v \rangle (n_{S, \rm{eq}}^2 - n_S^2) - \Gamma_S n_S,
\end{equation}
where \(n_{S, \rm{eq}}\) is the number density in thermal equilibrium, and \(\Gamma_S\) is the decay rate of \(S\). The energy density \(\rho_S = \langle E_S \rangle n_S\), where the average energy per particle is approximately \(\langle E_S \rangle \sim \sqrt{m_S^2 + 9T^2}\) and $T$ is the visible sector temperature, is initially insignificant. \\

The intermediate energy density of \(S\) can be set by either freeze-out or freeze-in mechanisms, depending on the interaction rate \(\langle \sigma_S v\rangle\). This decoupling leaves behind a residual energy density that redshifts as matter once \(S\) becomes non-relativistic. This energy density can surpass that of radiation, leading to \(S\) dominating if \(\rho_S^{\rm FO/FI}(T) > \rho_R(T)\). For both decoupling mechanisms, we need to calculate $H_{\rm EMD}$, as the times at which matter domination starts and ends will be the observables imprinted on the GW background by CS. For the freeze-out (FO) scenario, the redshifted frozen number density of $S$ particles at the beginning of the EMD is:
\begin{equation}
n_{S,\rm EMD} = n_{S, \rm FO} \left( \frac{H_{\rm EMD}}{H_{\rm FO}}\right)^{3/2} = \frac{H_{\rm EMD}^{3/2}}{\langle \sigma_S v \rangle H_{\rm FO}^{1/2}},
\end{equation}
with the number density at FO, $n_{S,\rm FO}={H_{\rm FO}}/{\langle \sigma_S v\rangle} $. Since at the onset of EMD: $\rho_{S,\rm EMD}\sim \rho_{R, \rm EMD}\sim 3 H_{\rm EMD}^2 M_{\rm Pl}^2$, one can simplify 
\begin{equation}\label{hemd}
    H_{\rm EMD}\simeq \frac{m_S^2}{9\langle \sigma_S v\rangle^2 M_{\rm Pl}^4 H_{\rm FO}}.
\end{equation}

When the dark scalar $S$ undergoes freeze-out while still relativistic, the annihilation rate must be sufficiently high for $S$ to reach equilibrium in its relativistic state. However, the rate should not be so high that $S$ remains in equilibrium after becoming non-relativistic. This process, known as relativistic freeze-out (RFO), results in the longest possible early matter-dominated (EMD) period. In this scenario, we have:
\begin{equation}
    H_{\rm EMD, RFO}\simeq \frac{30\sqrt{10} \zeta(3)^2 m_S^2  }{\pi^7 g_*^{3/2} M_{\rm Pl}}.
\end{equation}
If instead, the annihilation rate of $S$ is sufficiently high to keep it in equilibrium with the standard thermal bath below 
$T\sim m_S$, decoupling will occur through non-relativistic freeze-out (NRFO). This leads to a smaller frozen number density and, consequently, a later onset of the EMD era, characterized by \cite{Allahverdi:2021grt,Allahverdi:2022zqr}:
\begin{equation}
    H_{\rm EMD, NRFO}\simeq \frac{\sqrt{5}(\mathcal{W}\left(P\right))^2}{6\pi \sqrt{2 g_*} M_{\rm Pl}^3 }
\end{equation}
with $P={-8g_{*} \pi^5}/{45 \langle\sigma_S v\rangle^2 g_S^2 M_{\rm Pl}^2 m_S^2}$ and $\mathcal{W}(P)$ a Lambert function. Conversely, if the annihilation rate is lower than that required for RFO, $S$ will not achieve local chemical and thermal equilibrium. Instead, decoupling will occur via freeze-in (FI), resulting in a smaller out-of-equilibrium number density and thereby shortening the duration of the EMD with
\begin{equation}
    H_{\rm EMD,FI}\simeq \frac{90^3\zeta(3)^4g_S^4M_{\rm Pl}^2 \langle\sigma_S v\rangle^2 m_S^2 H_i }{9\pi^{14} g_*^3}
\end{equation}
where $H_i$ is the initial Hubble scale. Therefore, based on the interaction rate and the condition \(H_{\rm EMD} > \Gamma_S\) indicating a finite duration of EMD, it is possible to differentiate between the various decoupling mechanisms. To have a quantitative understanding of the duration of EMD we introduce a parameter $N_e$, which is defined as,
\begin{equation}
    e^{N_e}=\left(\frac{a_{\rm dom}}{a_{\rm dec}}\right)=\left(\frac{t_{\rm dec}}{t_{\rm dom}}\right)^{2/3}=\left(\frac{H_{\rm EMD}}{\Gamma_{S}}\right)^{2/3}
\end{equation}

\medskip
\section{Cosmic strings as a cosmic witness}\label{sec:strings}

Any spontaneous breaking of $U(1)$ symmetry in the early universe gives rise to a network of \emph{cosmic strings} (CS), a kind of topological defect configuration. Depending upon the $U(1)$ symmetry being \emph{local} or \emph{global}, such CS networks have different properties and evolve differently. For original work on this topic see Refs. by Kibble \cite{Kibble:1976sj}, and Vilenkin's textbook \cite{Vilenkin:2000jqa} which provide a very comprehensive overview. For discussions regarding the production of gravitational radiation from cosmic strings, see reviews in Refs. \cite{Gouttenoire:2019kij, Auclair:2019wcv, Gouttenoire:2022gwi, Simakachorn:2022yjy}.

To be precise, CS are typical field configurations at the top of any Mexican hat potential of the scalar which breaks $U(1)$. The symmetry breaking energy scale (its inverse) basically determines the size of the core within which the energy associated with the field is concentrated. This is quite smaller than the size of the cosmological horizon. therefore, CS can be treated as infinitely thin classical objects something also known as the \emph{Nambu-Goto} approximation. According to this, it carries an energy per unit length denoted by
\begin{equation}
\mu = 2\pi n \, v_{\rm CS}^2 \times \begin{cases}
1 ~ ~ ~ & {\rm for ~ local ~ strings},\\
\log(v_{\rm CS} t)  ~ ~ ~ & {\rm for ~ global ~ strings}.
\end{cases}
\label{string_tension}
\end{equation} 
Here $v_{\rm CS}$ (or $v_{\rm BL}$) represents the vacuum expectation value (VEV) of the scalar field which is responsible for the CS, and $n$ denotes the winding number, which we set to ($n=1$) since this is the only known stable configuration \cite{Laguna:1989hn}. We remark that for the scenario involving the global cosmic strings, there arises a logarithmic dependence with respect to the VEV due to the presence of massless Goldstone mode. The Goldstone bosons lead to typical long-range gradient energy \cite{Vilenkin:2000jqa}. The temperature at which CS network form can be written as 
\begin{align}
T_\textrm{form} &\simeq v_{\rm CS} \simeq 10^{11} \textrm{ GeV}\left(\frac{G\mu}{10^{-15}}\right)^{1/2},
\label{string_formation_cutoff}
\end{align}
where the second equality denotes the case for local strings.

 The characteristic length of long strings also known as the correlation length $L$, is related to the string tension via the following relation, $L =\sqrt{\mu/\rho_\infty}$, where $\rho_\infty$ represents the energy density of the long strings. After the formation of the cosmic string network, it interacts strongly with the thermal plasma leading to a damped kind of motion. Once the damping phase concludes, the strings oscillate and enter a scaling evolution phase characterized by two
competing dynamics: the Hubble expansion which stretches the correlation length of the strings and the fragmentation of such long strings into closed loops. These closed loops then oscillate independently afterwards, generating gravitational radiation as GWs and particle radiation as particle production \cite{Vachaspati:1984gt, Allen:1991bk,Vilenkin:2000jqa}\footnote{There is an ongoing controversy concerning the precise predictions of field theoretic methods and string simulations on the lattice, see Refs. \cite{Vincent:1997cx,Matsunami:2019fss,Hindmarsh:2021mnl,Blanco-Pillado:2023sap} for such details}.
In spite of these two competing dynamics lies an attractor solution known as the scaling regime, in which the characteristic length $L$ scales as cosmic time $t$. In this scaling regime, for a constant string tension, the energy density evolves as $\rho_\infty \propto t^{-2}$. Consequently, the cosmic string network becomes the same as any cosmological background energy density $\rho_{\rm bkg} \propto t^{-2}$ with the constant of proportionality given by the small quantity $G\sim M_{\rm Pl}^{-2}$, where $G$ is the Newton constant. Unlike topological defects like domain walls or other cosmic relics like dark matter, this particular scaling behaviour of the cosmic string network avoids cosmic sting domination over the Universe's energy density budget. We remark here the difference between local and global cosmic strings: the local string loops decay into GWs after $0.001/G\mu \gg 1$ Hubble times, while global string loops decay into Goldstone modes in less than one Hubble time.

So to summarise there are two steps in the process of emission of GW: first, the loops are created at time $t_i$ with string loop size parameter denoted by $\alpha$. The rate of loop formation rate can be expressed as \cite{Vilenkin:2000jqa}:

\begin{align}
\frac{dn_{\rm loop}}{dt_i} = 0.1 \frac{C_{\rm eff}(t_i)}{\alpha t_i^4}.
\end{align}
Here the factor $0.1$ indicates that $90\%$ of the loop population consists of small, highly-boosted loops. Consequently, these are usually red-shifted away without contributing to the GW radiation in any significant manner\cite{Blanco-Pillado:2013qja}. The loop formation efficiency factor $C_{\rm eff}$, depicting the decay rate at which local strings attain the asymptotic values or scaling solutions is given by $C_{\rm eff} \simeq 0.39$ and $5.4$ during the matter-dominated (MD) and radiation-dominated (RD) epochs respectively \cite{Gouttenoire:2019kij}. 
For global strings, besides the loop formation, the long strings also lose energy into light Goldstone boson particle production, due to which the loop production efficiency becomes logarithmically suppressed. Analytically, this can be understood as  $C_{\rm eff} \sim \mathcal{O}(1)$ across all cosmological epochs \cite{Chang:2019mza, Gouttenoire:2019kij,Chang:2021afa,Ramberg:2019dgi}. 
 In order to accurately determine $C_{\rm eff}(t)$ one needs to solve the velocity-dependent one-scale (VOS) equations which govern the evolution of the string network. However, in this study, we employ simplistic scaling solutions instead, as they effectively capture the essential qualitative features of the model that we need to demonstrate for the complementarity studies that we allude to, in this paper.

In the second step, the string loops oscillate emitting GW radiation. Numerical simulations \cite{Blanco-Pillado:2013qja} indicate that the GW spectrum is primarily generated due to loops of the largest size, which correspond to about $10\%$ of the horizon. Adopting a monochromatic probability distribution for such loop sizes 
\begin{align}
\mathcal{P}_{\rm loop}(\alpha) = \delta(\alpha - 0.1)
\end{align}
and since their formation at $\tilde t > t_i$, loops of length $l(\tilde{t})$ oscillate and radiate a discrete spectrum of GWs with frequencies set by the relation
\begin{equation}
\label{eq:emitted_frequency}
\tilde{f} = 2k/l(\tilde{t}), \qquad  k \in \mathbb{Z}^+.
\end{equation}
The present-day red-shifted frequency is given by $f=\tilde{f} a(\tilde{t})/a_0$.
Please note that the GW emission power from a loop is quite independent of the size of the loop \cite{Gouttenoire:2019kij}. Let us look into each Fourier mode $k$, which is given by,
\begin{align}
P_{\rm GW}^{(k)} = \Gamma^{(k)} G\mu^2, \qquad \text{with} \quad \Gamma^{(k)} = \frac{\Gamma k^{-{\delta}}}{\sum_{p=1}^\infty p^{-\delta}},
\label{eq:power_emisssion_GW_strings}
\end{align}
and $\Gamma = 50$ for local cosmic strings \cite{Blanco-Pillado:2017oxo} and global strings \cite{Gorghetto:2021fsn}. 
The value of $\delta$ depicts the scenario if this depends on whether high Fourier modes are dominated by cusps (for which $\delta = 4/3$), kinks (for which $\delta =5/3$), or kink-kink collisions  (for which $\delta=2$) \cite{Olmez:2010bi}. This value greatly influences the frequency slope at the transition between early matter domination and radiation domination. Assuming $\delta = 4/3$, that is the small-scale case to be dominated by cusps, we will proceed with the estimations.
Since the loops continuously lose energy via the emission of GWs or via Goldstone boson production, their length $l$ keeps on shrinking as 
\begin{align}
l(\tilde{t}) =  \alpha t_i - (\Gamma G \mu + \kappa) (\tilde{t} - t_i). \label{eq:length_shrink}
\end{align}
where $\Gamma G \mu$ and $\kappa$ are the shrinking rates associated with the emission of GW background and particle productions, respectively.
Local-string loops primarily decay via gravitational radiation emission ($\kappa = 0$), whereas global string loops predominantly decay into Goldstone bosons efficiently, characterised by $\kappa = \Gamma_{\rm Gold}/2 \pi \log(v_{\rm CS} t) \gg \Gamma G\mu$, where $\Gamma_{\rm Gold} \simeq 65 $ \cite{Vilenkin:1986ku}. 

Finally, the total GW energy density is estimated by summing
over all the $k$ modes leading to	\begin{align}
\Omega_{\rm GW}(f) &=\sum_k\frac{1}{\rho_c}\cdot\frac{2k}{f}\cdot\frac{\mathcal{F}_\alpha \,\Gamma^{(k)}G\mu^2}{\alpha(\alpha+\Gamma G \mu + \kappa)} \times \nonumber\\
&\hspace{3em}  \int^{t_0}_{t_{\rm osc}}d\tilde{t} \, \frac{C_{\rm{eff}}(t_i)}{t_i^4}\left[\frac{a(\tilde{t})}{a(t_0)}\right]^5\left[\frac{a(t_i)}{a(\tilde{t})}\right]^3\Theta\left(t_i-\frac{l_*}{\alpha}\right)\Theta(t_i-t_{\rm osc}).
	\label{eq:master_eq_ready_to_use}
	\end{align}
Here $\rho_c$ denotes the critical energy density of the universe, and $\mathcal{F}_\alpha \simeq 0.1$ is an efficiency factor. Please note that the integral in Eq. \eqref{eq:master_eq_ready_to_use} is constrained by two Heaviside functions, $\Theta\left(t_i-\frac{l_*}{\alpha}\right)\Theta(t_i-t_{\rm osc})$,
which acts as a cut-off at a very high frequency $f_*$, beyond
which the GW spectrum decreases as $f^{-1/3}$ slope when summed over a large number of modes. The quantity $t_{\rm osc}=\text{Max}\left[t_{\rm form},\,t_{\rm fric} \right]$ indicates the time until the motion of the string network is dampened by friction or until loops that could have formed prior to the formation of the network are eliminated. Additionally, $l_*$ signifies a critical length above which GW emission predominates over particle production, as demonstrated by high-resolution numerical simulations. The Eq.\eqref{eq:master_eq_ready_to_use} is applicable to both local and global strings after applying Eqs.~\eqref{eq:emitted_frequency} along with the selection of an appropriate value of $\kappa$.

At high frequencies, under standard cosmological evolution, the GW spectrum emitted by local strings is nearly flat with an amplitude of order
\begin{equation}
    \Omega_{\rm std}^{\rm Local} h^2 \simeq 15 \pi\Omega_r h^2 \, \Delta_T\, C_{\rm eff}^{\rm rad,l} \mathcal{F}_\alpha\left(\frac{\alpha G \mu}{\Gamma}\right)^{1/2},
\end{equation}
where $\Omega_{r}h^2 \simeq 4.2 \times 10^{-5}$\cite{ParticleDataGroup:2020ssz} and deviations from flatness may occur due to variations in the number of relativistic degrees of freedom accounted for in \cite{Gouttenoire:2019kij} 
\begin{align}
\label{eq:Delta_R}
\Delta_T \equiv \left( \frac{g_*(T)}{g_*(T_0)}\right)\left(\frac{g_{*s}(T_0)}{g_{*s}(T)} \right)^{4/3} .
\end{align}
In contrast, the GW spectrum produced by global strings is suppressed at higher frequencies and can be approximated as \cite{Gouttenoire:2019kij}
\begin{equation}
    \Omega_{\rm std}^{\rm Global} h^2 \sim  90\Omega_r h^2 \, \Delta_T\,C_{\rm eff}^{\rm rad,g} \mathcal{F}_\alpha\left(\frac{\Gamma}{\Gamma_{\rm gold}}\right) \left(\frac{v_{\rm CS}}{M_{\rm pl}}\right)^{4} \log^3\left( v_{\rm CS} \tilde{t}_{\rm M}\right),
\end{equation}
where the time of maximum emission $\tilde{t}_M$ is given as
\begin{equation}
    \tilde{t}_M =\frac{1}{t_0} \frac{4}{\alpha^2} \left( \frac{1}{f}\right)^2 \left( \frac{\alpha+\Gamma G \mu +\kappa}{\Gamma G \mu +\kappa}\right)^2.
\end{equation}
For a comprehensive discussion on the differences between GW from local and global strings, one can follow ~\cite{Gouttenoire:2019kij,Ghoshal:2023sfa}.

Following Refs \cite{Ghoshal:2023sfa}, in scenarios featuring an early matter-dominated epoch instead, the plateau takes a spectral turnover at a high frequency $f_{\rm brk}$, beyond which the spectrum falls as $\Omega_{\rm GW}(f>f_{\rm brk})\propto f^{-1/3}$ for cusp-dominated structures when summing over a large number of modes. This spectral break frequency $f_{\rm brk}$ for both string classes can be estimated as,
\begin{equation}
    f_{\rm brk}^{\rm local}=6.32\times 10^{-3}\text{Hz} \left( \frac{T_{\rm brk}}{\rm GeV}\right) \left( \frac{0.1\times 50 \times 10^{-12}}{\alpha \Gamma G \mu}\right)^{1/2}  \left( \frac{g_{*}(T_{\rm brk})}{g_{*}(T_0)}\right)^{1/4}
\end{equation}
and
\begin{equation} \label{turning_point_CS}
    f_{\rm brk}^{\rm global}= 8.9\times 10^{-7}\text{Hz} \left( \frac{T_{\rm brk}}{\rm GeV}\right) \left( \frac{0.1}{\alpha}\right)  \left( \frac{g_{*}(T_{\rm brk})}{g_{*}(T_0)}\right)^{1/4}
\end{equation}
where $T_{\rm brk}\equiv T_{\rm end}$ is the temperature corresponding to the end of early matter domination. It is important to note that while $f_{\rm brk}^{\rm global}$ has a linear dependence on $T_{\rm brk}$, it shows insensitivity to the symmetry breaking scale $v_{\rm CS}$, which is different from what we observe with local strings. Such a situation is clearly depicted in the GW spectrum shown in Fig. \ref{fig7}.

\medskip

\section{Lab searches for long-lived particles}\label{sec:longlived}

In recent years there have been intensive efforts to constrain the possible existence of light
mediators connecting the SM to a hidden sector, the scalar singlet with the Higgs portal being
a prime example. The parameter space of singlet mass $m_S$ and mixing $\theta$ (the singlet-Higgs mixing angle)
is constrained by a variety of beam-dump, collider and rare decay experiments, and by
cosmology (Big Bang nucleosynthesis), astrophysics (supernova cooling) and dark matter
direct searches. A large region of parameter space with $\theta \lesssim 10^{-3}$ and $m_S \gtrsim 0.1 $ GeV remains open, and parts of this will be targeted by the upcoming SHiP experiment \cite{SHiP:2015vad}.

In this section we investigate the prospects of searching for this light scalar in intensity frontier and lifetime frontier experiments which look for light, weakly
interacting, electrically neutral long-lived particles. These experiments are capable of probing up to extremely small mixing angles or couplings. Dark sectors with
light degrees of freedom can be searched for, with a variety of
experiments at the luminosity frontier, including proton~\cite{Batell:2009di, deNiverville:2011it, deNiverville:2012ij, Kahn:2014sra, LBNE:2013dhi, Soper:2014ska, Dobrescu:2014ita, Coloma:2015pih, deNiverville:2016rqh, MiniBooNE:2017nqe, MiniBooNEDM:2018cxm, MATHUSLA:2018bqv, FASER:2018bac}, electron~\cite{Bjorken:2009mm, Izaguirre:2013uxa, Diamond:2013oda, Izaguirre:2014dua, Batell:2014mga, BaBar:2017tiz, Berlin:2018bsc, Banerjee:2019pds} and positron fixed target facilities that exist and are being developed worldwide~\cite{Accardi:2020swt}. Here we show, the allowed target parameter space of such dark sector lies well within the reach of several such experiments given the mass and mixing of the light scalar\footnote{Light dark sector scenarios have been explored in great detail over the past few years in recent times~\cite{Bramante:2016yju, Alexander:2016aln, Battaglieri:2017aum, Darme:2017glc, Winkler:2018qyg, Egana-Ugrinovic:2019wzj, Okada:2019opp, Foroughi-Abari:2020gju, Agrawal:2021dbo,Nardi:2018cxi}.}. Such a tiny mixing angle is indeed within the reach of upcoming and current experiments like DUNE~\cite{DUNE:2015lol, Berryman:2019dme},  FASER-II~\cite{Feng:2017vli, FASER:2018eoc, FASER:2018bac, FASER:2019aik}, PS191~\cite{Bernardi:1985ny, Gorbunov:2021ccu}, DarkQuest-Phase2~\cite{Batell:2020vqn}, MATHUSLA~\cite{Curtin:2018mvb} and SHiP~\cite{SHiP:2015vad}.\\

To give an example, the ForwArd Search ExpeRiment (FASER) \cite{Feng:2017vli, FASER:2018eoc, FASER:2018bac, FASER:2019aik} searches for light, weakly
interacting, electrically neutral long-lived particles which may be produced at the Large Hadron Collider (LHC) in CERN. In the experiment, such a particle detector is located along the proton beam trajectory about 480 meters downstream from the interaction point within the ATLAS detector at the LHC. This particular experimental setup is dedicated to the search for light, long-lived particles due to the following well-motivated reasons: (i) the High-Luminosity upgrade of the LHC (also known as HL-LHC) will be able to produce a huge number of hadronic particles in the forward region, which are likely to decay into such light long-lived particles if they exist. Now because of this huge flux of such particles, even if its decay is an extremely rare event, the final resultant event that one may be able to achieve in this experiment is actually quite sizable for such long-lived particle production; (ii) these light particles which are produced are highly boosted in
the direction of the beam and will end up in the forward region (iii) Due to very weak interactions involved, such particles may have a very large decay length, of $\mathcal{O}$( 100 m) which means they travel about 100 meters before they decay. The displaced vertex signature from such decays is therefore completely devoid of SM background, which stops after about 1 cm or so. Consequently, this becomes almost a background-free search.

In our case, the long-lived scalar $S$ decays into light SM fermions via the small mixing with the SM Higgs, whose decay width can be given as,
\begin{equation}
    \Gamma(S\rightarrow f \bar{f}) \simeq  \Gamma_{\rm SM}^{\rm total}(m_S) \sin^2 \theta,
\end{equation}
with,
\begin{equation}
    \Gamma_{\rm SM}^{\rm total}(m_S)= \Gamma_{H\rightarrow \ell \bar{\ell}}+ \Gamma_{H\rightarrow q \bar{q}}+ \Gamma_{H\rightarrow W^{-}W^{+}} + \Gamma_{H\rightarrow ZZ}+\Gamma_{H\rightarrow K K}+\Gamma_{H\rightarrow \pi \pi }
\end{equation}
where \cite{Fradette:2017sdd, Okada:2019opp},
\begin{equation} \label{partial widths}
\begin{aligned}
    {} & \Gamma_{H\rightarrow \ell \bar{\ell}}=\frac{m_S}{8\pi}\left( \frac{m_\ell}{v}\right)^2 \left( 1-\frac{4 m_\ell^2}{m_S^2}\right)^{3/2},\\
    & \Gamma_{H\rightarrow q \bar{q}} = 3 \Gamma_{H\rightarrow \ell \bar{\ell}} \bigg|_{\ell\rightarrow q} \left[ 1+\frac{4\alpha_{\rm QCD}}{3\pi}\left( \frac{9}{4}+\frac{3}{2}\log\frac{m_q^2}{m_S^2}\right)\right],\\
    & \Gamma_{H\rightarrow W^{-}W^{+}} =\frac{m_S^3}{16\pi v^2}\left(1- \frac{4m_W^2}{m_S^2} \right)^{1/2} \left( 1- \frac{4m_W^2}{m_S^2}+ \frac{12 m_W^4}{m_S^4}\right)  ,\\
    & \Gamma_{H\rightarrow ZZ}=  \frac{m_S^3}{32\pi v^2}\left(1- \frac{4m_Z^2}{m_S^2} \right)^{1/2} \left( 1- \frac{4m_Z^2}{m_S^2}+ \frac{12 m_Z^4}{m_S^4}\right) , \\
    &  \Gamma_{H\rightarrow \pi \pi}=\frac{m_S^3}{16\pi v^2}\left(\frac{2}{9}+ \frac{11 m_\pi^2}{m_S^2} \right)^2 \left(1- \frac{4m_\pi^2}{m_S^2} \right)^{1/2} ,\\
    & \Gamma_{H\rightarrow K K}=\frac{m_S^3}{8\pi v^2}\frac{27}{13}\left( 1- \frac{4 m_K^2}{m_S^2}\right)^{3/2},
    \end{aligned}
\end{equation}
where $\Gamma_{\rm SM}^{\rm total}(m_S)$ denotes the total decay width of the SM Higgs boson, restricted to channels that are kinematically open for the scalar decay and rescaled by the factor $m_S/m_H$ and we have taken the QCD running coupling constant, denoted as $\alpha_{\rm QCD}$, to be approximately $\alpha_{\rm QCD}(M_Z) \sim 0.1181$\footnote{Although, due to asymptotic freedom, $\alpha_{\rm QCD}$ decreases at high energies (short distances) and increases at low energies (long distances), we don't expect much deviation in our results given the broad energy spectrum of interest for gravitational wave signatures.}. In the following sections, we explore how the decay width, proportional to $\sin^2 \theta$, can be very small, and as a result, the scalar can be quite long-lived to have the potential to imprint its characteristic signatures in lab searches.


\medskip

\section{Complementarity between GW and laboratory searches}\label{sec:comp}



\medskip
Probing the aforementioned long-lived scalar \(S\) in upcoming colliders is very challenging. However, in the early universe, non-negligible processes such as inflaton decay or moduli decay can result in a significant initial abundance of \(S\), potentially altering the standard scenario. For freeze-out, a sufficiently high interaction rate would eliminate any excess initial abundance by maintaining equilibrium. In contrast, for freeze-in, due to the weak interaction strength, it is valuable to investigate how such a small initial abundance can extend the duration of the EMD period. Assuming an initial temperature at $T_i$, let's consider a scenario where species $S$ has a non-zero initial abundance, denoted by $\eta_S$, which is proportional to the ratio of its energy density $\rho_S(T_i)$ to that of the radiation component $\rho_R(T_i)$ at that temperature. Furthermore, we'll simplify our analysis by assuming that, due to their extremely weak interactions, these particles are initially relativistic but become non-relativistic as their mass approaches the temperature scale, so that one can redshift the initial abundance to get the abundance at the onset of EMD era,
\begin{equation}
    \frac{n_{S,\rm EMD}}{n_{S,i}}=\left( \frac{H_{\rm EMD}}{H_i}\right)^{3/2}.
\end{equation}
By rearranging this equation, we can obtain a rough estimate of the Hubble scale during the EMD era, which is given by:
\begin{equation}
    H_{\rm EMD}\sim \eta_{S}^2 H_i.
\end{equation}
We discuss the impact of such EMD on the complementarity in the subsequent sections.
\begin{figure}
\centering
\includegraphics[scale=.5]{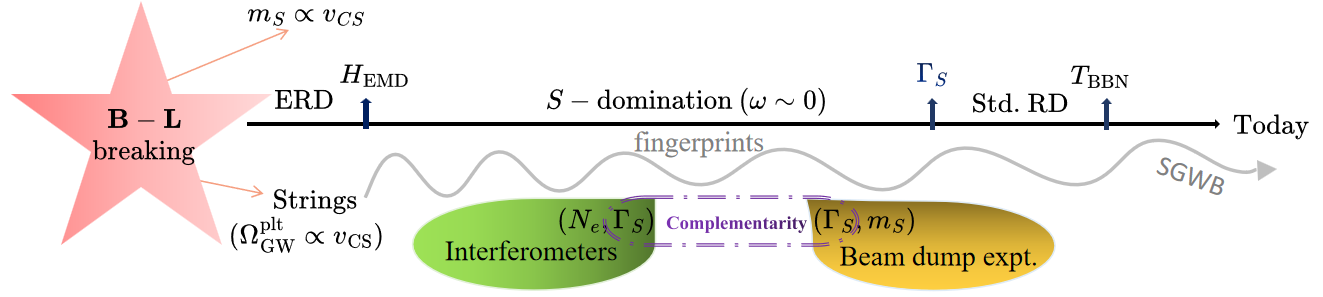} 
 	\caption{\it Schematic timeline of key events: Following inflation, a $(B-L)$ phase transition generates cosmic string networks during the early radiation-dominated (ERD) era, which become a dominant source of GWs primarily through loop production. Simultaneously, a real scalar singlet, $S$, acquires a mass dependent on the $(B-L)$ Higgs vacuum expectation value. Starting from $H_{\rm EMD}$ to $\Gamma_S$, this scalar dominates the universe's energy density, after which the universe transitions to standard radiation domination. The long-lived scalar leaves an imprint on the high-amplitude GWs generated by cosmic strings, with the duration ($N_e$) and end of matter domination ($\Gamma_S$) influencing the spectral shapes, which would otherwise produce a scale-invariant, flat plateau at high frequencies. Upcoming beam dump experiments offer a compelling and complementary approach to search for and constrain such long-lived BSM particles based on their mass ($m_S$) and decay width ($\Gamma_S$) to the SM states.}\label{fig1}
 \end{figure}








\medskip

\section{Results}

\begin{table}[]
\begin{tabular}{|llll|}
\hline
\multicolumn{4}{|l|}{Freeze-in: $H=10^{12}{\:\rm GeV}, \eta_S=10^{-13}$}                                         \\ \hline
\multicolumn{1}{|l|}{Parameters} & \multicolumn{1}{l|}{$\sin \theta$} & \multicolumn{1}{l|}{$m_S$ (GeV)} & $N_e$ \\ \hline
\multicolumn{1}{|l|}{BP1}        & \multicolumn{1}{l|}{$10^{-5.38}$}  & \multicolumn{1}{l|}{$10^{0.13}$} & 4.15  \\ \hline
\multicolumn{1}{|l|}{BP2}        & \multicolumn{1}{l|}{$10^{-5.61}$}  & \multicolumn{1}{l|}{$10^{0.01}$} & 6.8   \\ \hline
\multicolumn{1}{|l|}{BP3}        & \multicolumn{1}{l|}{$10^{-5.45}$}  & \multicolumn{1}{l|}{$10^{-0.5}$} & 8.05  \\ \hline
\end{tabular}
\caption{\it The three benchmark points considered in this paper for the modified freeze-in scenario. In each case, there is a possibility of complementary detection by lab searches and GW observations.  }\label{t1}
\end{table}
In this section, we present our numerical results that describe decoupling through both standard freeze-out and freeze-in mechanisms. Additionally, we explore a modified, highly testable freeze-in setup. Finally, we quantitatively assess their complementary detection capabilities for future lab searches and GW observations. Specifically, an early matter-dominated (EMD) era, driven by a long-lived scalar field $S$ with a non-standard EOS ($\omega\sim0$), can imprint signatures onto the SGWB originating from cosmic strings, encoding valuable information about the scalar field's properties and the underlying BSM physics, as sketched in the schematic in Fig. \ref{fig1}. 

\par 
As discussed in previous sections, cosmic string-induced GWs exhibit a nearly scale-invariant plateau across mid to high frequencies, resulting in a potentially detectable signal. Specifically, such an EMD epoch, driven by $S$, would imprint distinct features onto this plateau. The duration of this matter domination ($N_e$), governed by the scalar field's decoupling mechanism (e.g., freeze-in or freeze-out, characterized by the scattering cross-section ($\langle \sigma v \rangle$), would be encoded in the plateau's shape. Additionally, the end of the EMD ($\Gamma_S$), determined by the scalar mixing with SM particles, would manifest as the plateau's mid-frequency spectral break ($f_{\rm brk}$). Concurrently, upcoming laboratory experiments are reaching unprecedented sensitivities in their search for long-lived particles. These searches directly probe the mass of such particles, their decay widths or their strength of mixing with the SM states. This opens exciting avenues for synergy between GW detections and lab searches. The complementary information gathered from these two frontiers holds the potential to unveil the nature of these well-motivated BSM scalar fields. \par
\begin{figure}[H]
\centering
\includegraphics[scale=.7]{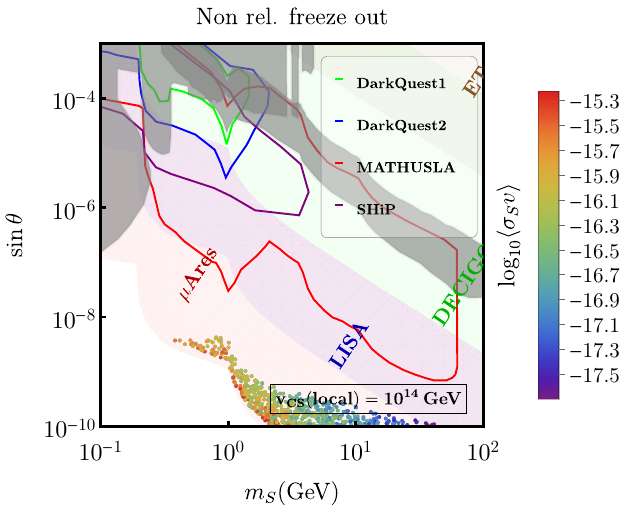} \includegraphics[scale=.7]{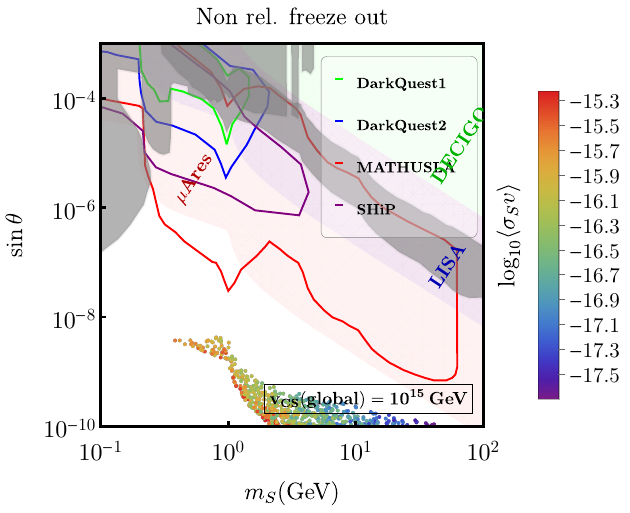}
\includegraphics[scale=.7]{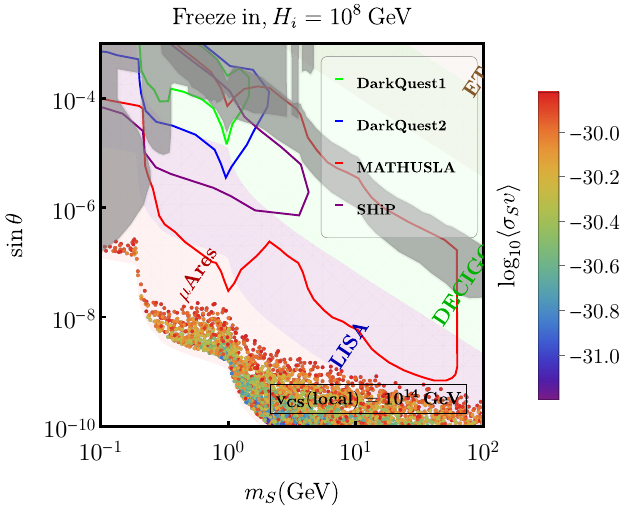} \includegraphics[scale=.7]{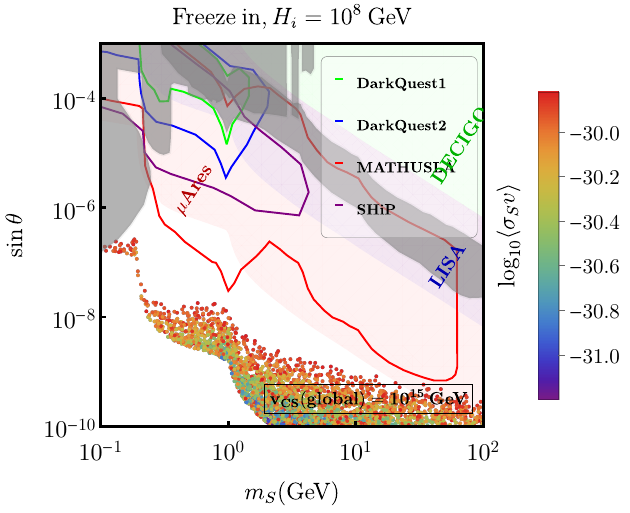}
\\
 	\caption{\it Complementarity of GW and laboratory searches (indicated by colored dots) for Top-left: standard non-relativistic FO with local $(B-L)$ strings with $v_{\rm CS}=10^{14}$ GeV, Top-right: standard non-relativistic FO with global $(B-L)$ strings with $v_{\rm CS}=10^{15}$ GeV, Bottom-left: standard FI with $H_i=10^8$ GeV and local $(B-L)$ strings with $v_{\rm CS}=10^{14}$ GeV, Bottom-right:  standard FI with $H_i=10^8$ GeV and global $(B-L)$ strings with $v_{\rm CS}=10^{15}$ GeV. The colored shaded regions represent the sensitivity curves of various GW experiments, such as $\mu$Ares (red), LISA (blue), DECIGO (green), and ET (brown), with the stipulation that the first turning point frequency to be detected with SNR $>10$. The gray shaded regions are excluded by CHARM, LHCb, CMS re-interpretation results. The solid lines show the sensitivity of upcoming long-lived particle searches like DarkQuest1, DarkQuest2, MATHUSLA, SHiP, etc. }\label{fig3}
 \end{figure}

 
Following this approach, we perform a comprehensive parameter scan considering both non-relativistic freeze-out and freeze-in scenarios. As depicted in Fig. \ref{fig3} and illustrated in Section \ref{sec:history}, in both cases, the number density at decoupling required for an early matter-dominated era exhibits a strong dependence on the scattering cross-section, $\langle \sigma v \rangle$. Our numerical analysis incorporates all possible decay channels of the scalar $S$ in the SM states. These decay modes are determined by the relation given in Eq.\eqref{partial widths}, which couples the scalar $S$ to the Higgs through a small mixing angle, $\theta$. From Fig.\ref{fig3} we observe that for non-relativistic freeze-out, a large scalar mixing angle leads to a high scattering cross-section (or an efficient annihilation), inhibiting a significant EMD. Conversely, for the freeze-in scenario, the weak coupling strength limits how long matter can dominate. Consequently, for both cases, nearly the entire portion of the parameter space remains inaccessible to upcoming long-lived particle searches. \par 
Decoupling through freeze-out occurs when $m_S\sim T$. For lighter scalar masses, this decoupling occurs at a much later time. Specifically, when the scalar mass $m_S\lesssim 10^{-0.5}$ GeV, decoupling happens around $T\sim T_{\rm BBN}$ which is incompatible with the observed abundances of primordial nucleons. For a fixed $m_S$, an increasing $\langle \sigma v\rangle$ implies a lower number density of $S$ during decoupling, leading to a shorter period of EMD. For very light $m_S$, it is not even possible to reach an EMD before BBN, corresponding to the lower cutoff in the $\sin\theta-m_S$ plot. In contrast, for low values of $\langle \sigma v \rangle$, the number density of $S$ during freeze-out will be large, which can lead to a significantly longer duration of EMD. It is possible to achieve an EMD even with a large mixing angle $\sin \theta$, which corresponds to a rapid decay mode. However, for very small values of $m_S$, the EMD begins close to BBN corresponding to the upper limit in the $\sin\theta-m_S$ plot. \par 
In the case of freeze-in for a fixed $m_S$, the number density of $S$ increases as $\langle \sigma v \rangle$ rises. Therefore, an EMD is possible even with a large mixing. On the other hand, a lower $\langle \sigma v \rangle$ leads to insufficient number density to initiate an EMD significantly earlier, which is constrained by BBN. This corresponds to the lower cutoff in the $\sin\theta-m_S$ plot. It is evident that, compared to a freeze-out scenario, a freeze-in scenario allows for the possibility of achieving an EMD even with masses as low as $m_S \sim 10^{-1}$ GeV. For a quantitative understanding of the detection criteria (after $t_{\rm obs}=4$ years) of GW signals, we employ a turning point sensitivity prescription, highlighting regions where a departure from the standard flat plateau, characterized by the turning point ($f_{\rm brk}$), could be detected with a signal-to-noise ratio (SNR) defined as \cite{Maggiore:1999vm, Schmitz:2020syl},
\begin{equation}
    \text{SNR}= \sqrt{t_{\rm obs} \int_{f_{\rm min}}^{f_{\rm max}}df \left( \frac{\Omega_{\rm GW}(f)}{\Omega_{\rm noise}(f)}\right)^2}>10,
\end{equation}
where $\Omega_{\rm noise}(f)$ is the noise curve of a given experiment and $f_{\rm min}$($f_{\rm max}$) are minimum (maximum)
 accessible frequency. Interestingly, the forthcoming GW detectors are sensitive enough to pinpoint any such imprints, as shown in all subplots of Fig.\ref{fig3} through color-shaded regions for $\mu$Ares \cite{Sesana:2019vho} (red), LISA \cite{LISA:2017pwj} (blue), DECIGO \cite{Kawamura:2020pcg} (green) and ET \cite{Punturo:2010zz} (brown) respectively, and thus may provide complementary information about the scalar mass and its interactions. An examination of the subplots reveals a significant difference in the sensitivities of GW detectors to EMD for local vs global symmetry breaking (cosmic strings). This discrepancy arises from the differing amplitudes of the corresponding cosmic string networks at high frequencies, as evident in Fig. \ref{fig7}.  \\

\begin{figure}[H]
\centering
\includegraphics[scale=.7]{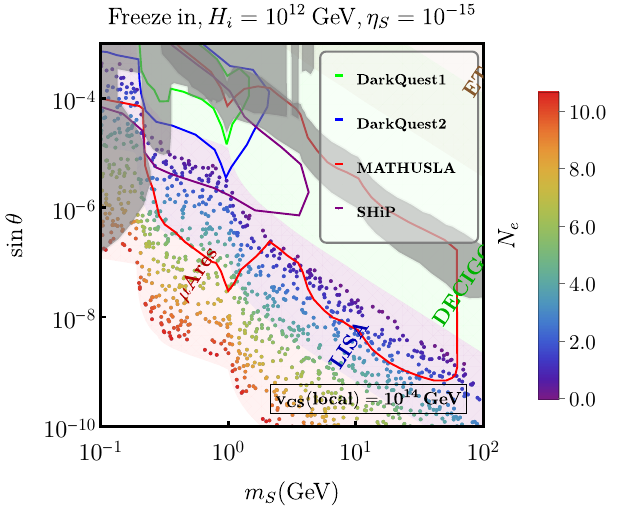} \includegraphics[scale=.7]{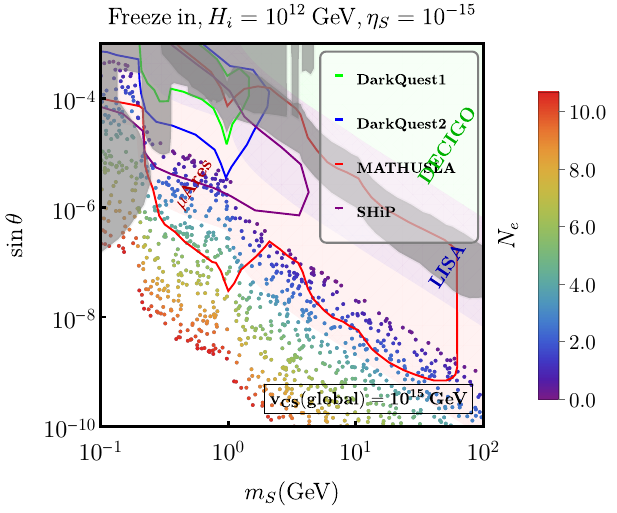}
\includegraphics[scale=.7]{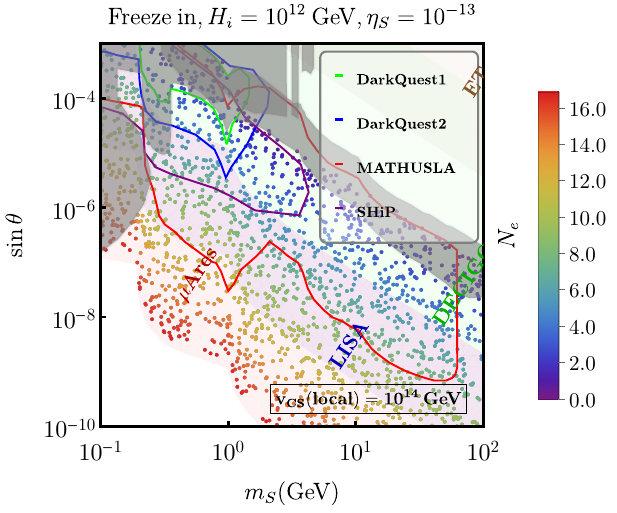} \includegraphics[scale=.7]{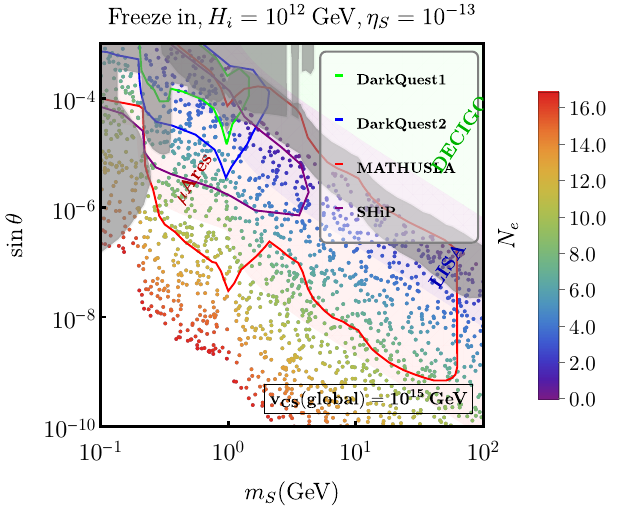}
 	\caption{\it Complementarity of GW and laboratory searches with a non-zero initial $S$- abundance(indicated by colored dots) for Top-left: freeze-in with local $(B-L)$ strings with $v_{\rm CS}=10^{14}$ GeV, initial Hubble scale $H_i=10^{12}$ GeV and $\eta_S=10^{-15}$, Top-right: same as the earlier with global $(B-L)$ strings with $v_{\rm CS}=10^{15}$ GeV, Bottom-left: freeze-in with local $(B-L)$ strings with $v_{\rm CS}=10^{14}$ GeV, initial Hubble scale $H_i=10^{12}$ GeV and $\eta_S=10^{-13}$, Bottom-right:  same as the earlier with global $(B-L)$ strings with $v_{\rm CS}=10^{15}$ GeV. As above, the shaded regions correspond to the ability of a GW detector to ``see'' the end of matter domination.}\label{fig4}
 \end{figure}
Even if any non-standard processes generate a significant abundance of the scalar field $S$ during the early universe's evolution, a large scattering cross-section in the freeze-out scenario would lead to efficient annihilation. This annihilation would erase any initial excess abundance, driving the scalar field towards its standard thermal abundance prior to freeze-out.  In contrast to the freeze-out scenario, the weak scattering inherent to freeze-in allows for a non-zero initial abundance of the scalar field $S$ to impact the cosmological evolution significantly. Even a small initial abundance can lead to an extended period of matter domination due to the differing scaling behaviour of the scalar field energy density compared to radiation. In this case, the start time of the EMD is directly proportional to $\eta_S^2$. Therefore, a slight increase in $\eta_S$ leads to a significantly larger $N_e$, due to the increased number density of $S$, which results in an earlier start time for the EMD. Now, this makes large mixing, which is of interest for lab searches, possible, although with a reduced $N_e$ since large mixing causes the EMD to end much earlier than BBN. \par
\begin{figure}[H]
\centering
\includegraphics[scale=.7]{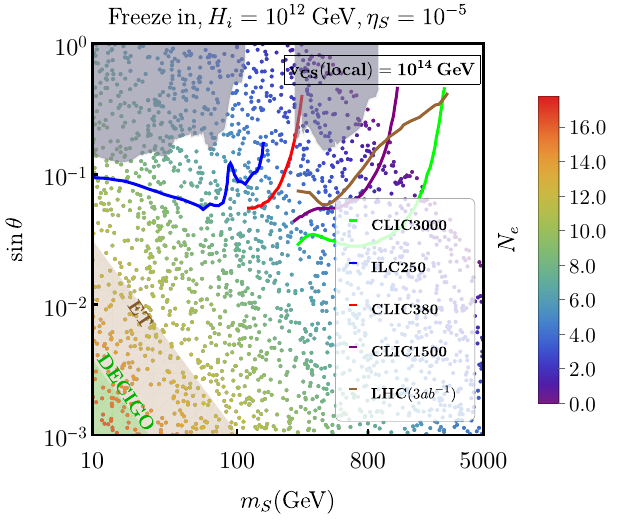} \includegraphics[scale=.7]{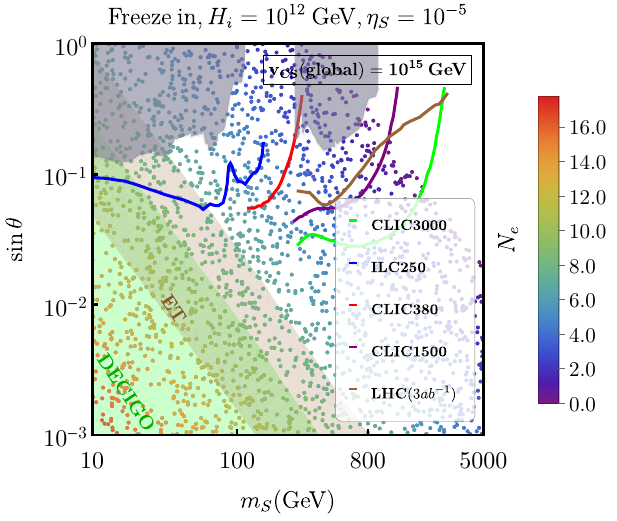}
 	\caption{\it Complementarity of GW and colliders dedicated for higher mass searches with a significantly larger initial $S$- abundance(indicated by colored dots) for Left: freeze-in with local $(B-L)$ strings with $v_{\rm CS}=10^{14}$ GeV, initial Hubble scale $H_i=10^{12}$ GeV and $\eta_S=10^{-5}$, Right: same as the earlier with global $(B-L)$ strings with $v_{\rm CS}=10^{15}$ GeV. As above, the shaded regions correspond to the ability of a GW detector to ``see'' the end of matter domination.}\label{fig5}
 \end{figure}
Similar to the previous scenarios, the lower bound in the $\sin\theta-m_S$ plot comes due to the requirement that the EMD should end before BBN. Fig. \ref{fig4} illustrates the impact of a non-zero initial abundance of the scalar field $S$ on the duration of matter domination. Even for a small initial abundance of $\eta_S=10^{-15}$, and $10^{-13}$, a substantial period of EMD can occur. However, this initial abundance effectively masks any information about the underlying scattering rate, rendering it challenging to extract from cosmological observables. Interestingly, this assumption of a non-zero initial abundance leads to an intriguing interplay between the sensitivities of future laboratory experiments and GW detectors. As demonstrated in Fig.\ref{fig4}, for both local and global cosmic strings, a significant portion of the viable parameter space becomes accessible to GW detectors and lab searches. Furthermore, for sufficiently large initial abundances as shown in Fig. \ref{fig5}, such as $\eta_S=10^{-5}$, even high-energy colliders may have the potential to probe the scalar field with masses at higher scales, further enhancing the synergy with GW observations.   \\

 \begin{figure}[H]
\centering
\includegraphics[scale=.7]{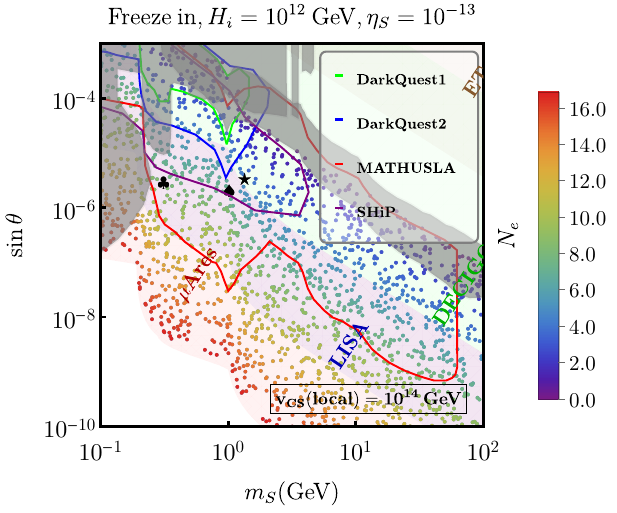} \includegraphics[scale=.7]{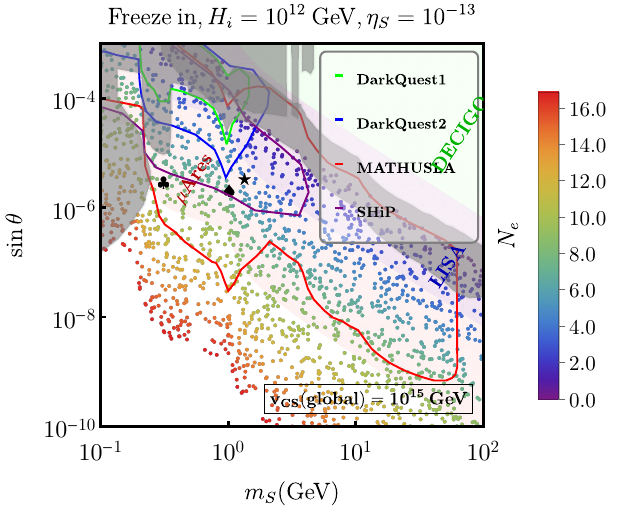}
 	\caption{\it Imprints of long-lived scalar $S$ on SGWB from cosmic strings and complementarity with collider searches for the benchmarks BP1 (``{\huge$\star$}"), BP2 ``$\spadesuit$"), BP3 ``$\clubsuit$") (see Table-\ref{t1}) shown in left: for local strings, right: for global strings. As above, for both plots, the shaded regions correspond to the ability of a GW detector to ``see'' the end of matter domination.}\label{fig6}
 \end{figure}

Fig. \ref{fig6} illustrates three benchmark cases (BP1, BP2, and BP3) detailed in Table \ref{t1} and represented by {\huge$\star$}, $\spadesuit$, and $\clubsuit$, respectively. All benchmark points fall well within the projected sensitivity ranges of upcoming long-lived particle searches like SHiP, MATHUSLA \cite{Curtin:2023skh} etc. The turning point frequencies for all benchmark points, for both local (with $v_{\rm CS}=10^{14}$ GeV) and global string (with $v_{\rm CS}=10^{15}$ GeV) cases, fall within the sensitivity range of current and planned GW searches. Fig.\ref{fig7} presents the SGWB spectra for both string classes, considering these three benchmarks and summing over a large number of modes, $k=10^3$. In principle, if one sums over an infinite number of modes there will be a characteristic  $f^{-1/3}$ fall after the break frequency ($f_{\rm brk}$) and a second plateau due to the loops formed at the early radiation epoch, for a finite duration of EMD. Although, a prolonged period of EMD started at a very early time ($T_{\rm dom}\gg T_{\rm dec}$) shifts the second kink ($f_{\rm dom}$) to much higher frequencies inaccessible to planned GW detectors, a short EMD may allow for the detection of this second spectral break. Interestingly, Such detection could provide valuable insights into the strength of EMD ($N_e$).



  \begin{figure}
\centering
\includegraphics[scale=.7]{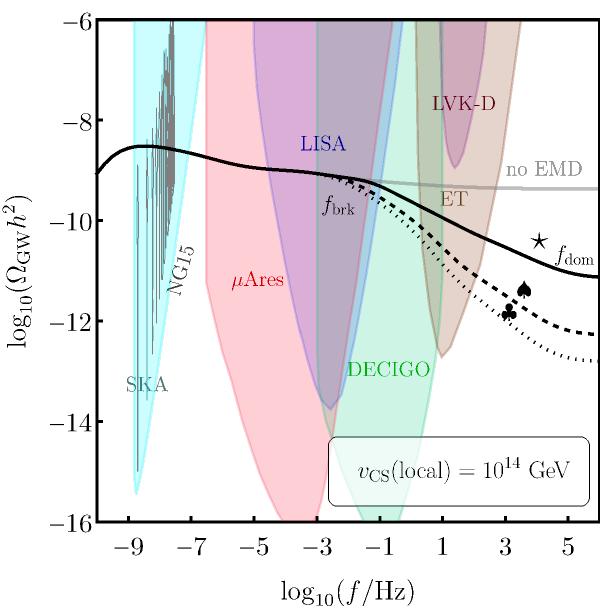}
\includegraphics[scale=.7]{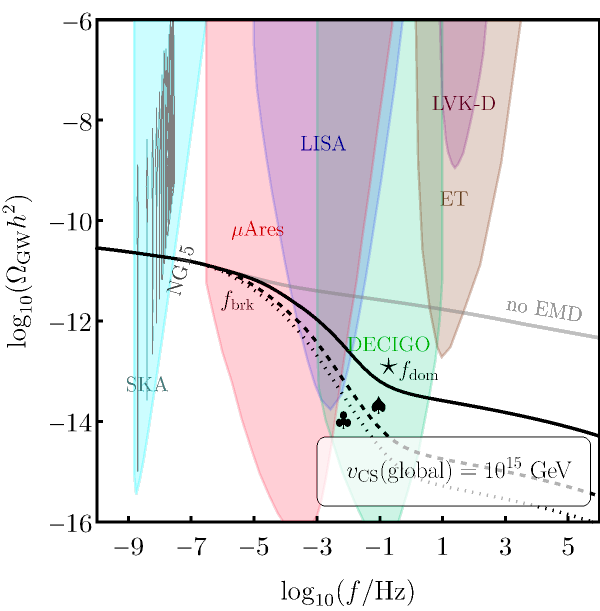}
 	\caption{\it The GW spectral shapes for local strings (left panel) and global strings (right panel) are shown with a period of early matter domination that lasts $N_e=$ $4.15$ (Solid), $6.8$ (dotted) and $8.05$ (dashed) respectively.  We also mark each benchmarks BP1 (``{\huge$\star$}"), BP2 ``$\spadesuit$"), BP3 ``$\clubsuit$") (see Table-\ref{t1}). The standard cosmological scenario without an EMD phase is shown as gray solid lines in both plots. In contrast, scenarios with an intermediate EMD phase exhibit a distinctive kink and a suppression in the gravitational wave spectrum, marking the period of matter domination, whereas the spectrum is relatively flat for frequencies that correspond to periods of radiation domination. Note that we show benchmarks where one timescale is observable as well as other benchmarks where one can observe both the start and end of matter domination.}\label{fig7}
 \end{figure}

\medskip

\section{Discussion and Conclusion}

Topological defects can be a unique probe into high-scale physics which is difficult to detect in terrestrial experiments \cite{Dasgupta:2022isg,Bhaumik:2022pil,Barman:2022yos,Ghoshal:2022jdt,Dunsky:2021tih,Bernal:2020ywq,Ghoshal:2020vud}. In this work, we demonstrate that they can also provide valuable insights into the physics of long-lived states, particularly due to a possible period of early matter domination. Specifically, we studied the thermal production mechanism of dark sectors, with cosmic strings providing a witness to any departure from the standard thermal history. As a special case we considered a scenario where the dark scalar is \textit{frozen in}, with a non-zero initial fraction of energy density, $\eta_S$, which can generally be zero. The suppressed decay rate of this scalar ensures its survival and contribution to a period of early matter domination, due to the oscillation of the dark sector scalar.

 The entire setup is therefore described by a minimal set of independent parameters: the mass and decay width of the scalar represented as, $m_S$ and $\Gamma_S$, and the energy fraction of $S$ particles at the end of inflation $\eta_S$. Assuming several instances of $U(1)$ symmetry breaking that lead to the VEV $\langle\Phi\rangle$ of scalar $\Phi$, which via a Higgs-portal like mixing $(\lambda_{\Phi S})$ leads to the generation of $S$ mass, we find a novel complementarity signature for long-lived $S$ in GW searches in LISA, ET, $\mu$Ares etc. We studied the effect of those different parameters on the shape of the GW spectrum with a period of early matter domination caused due to long-lived $S$. We showed that such a non-standard cosmological history leaves imprints on the GW signals, to be measured by upcoming detectors. We exhibited the GW spectral shapes that may be observed in LISA, ET, BBO, DECIGO and u-DECIGO.  

Because the production of the $S$ particle is intrinsic to our model and implicitly given by the lifetime, mass and decay branching fraction of $S$ into visible sectors, the interaction strength of the $S$ and SM particles (the Higgs for instance) can be simply extracted from each point in the parameter space. We showed that the long-lived particle $S$ can therefore be searched experimentally in light-dark sector searches involving intensity, lifetime, and beam dump experiments. We explored the parameter space in Fig.~\ref{fig6} and compared our results to the projected limits of laboratory searches such as FASER \cite{Feng:2017vli,FASER:2018eoc,FASER:2018bac,FASER:2019aik}, DUNE \cite{DUNE:2015lol,Berryman:2019dme}, DARKQUEST-2 \cite{Batell:2020vqn}, MATHUSLA \cite{Curtin:2018mvb}, PS191 \cite{Bernardi:1985ny,Gorbunov:2021ccu}, and SHIP \cite{SHiP:2015vad}. In particular, we found that an early matter-dominated period with equation-of-state parameter $\omega \approx 0$, with $m_S \sim \mathcal O(0.5-5)$ GeV,  could show detectable signals for DUNE, MATHUSLA, and SHIP. Interestingly, we also showed that this specific parameter region would yield detectable signals within four years of exposure for GW detectors such as LISA, ET and u-DECIGO, serving as a complementary smoking-gun signature. 

When studying the testability of our simple BSM model, we used a simple Higgs-portal setup and a heavy scalar particle. However, we insist on the fact that our prescription to search for complementary probes of new physics with laboratory and GW experiments is very general and can be applied to many other BSM scenarios that involve a non-standard cosmology \cite{Ghoshal:2021ief,Barman:2022njh,Banerjee:2022fiw,DEramo:2017gpl,Co:2015pka,Heurtier:2019eou, Heurtier:2017nwl,Heurtier:2022rhf} some of which include non-thermal DM production which again is otherwise very challenging to test in direct detection experiments due to feeble couplings to the visible sector. Following this prescription, we believe that many such realizations of gauged and globally extended $U(1)$ sectors in the early universe may lead to unique predictions of GW spectral shapes that can be detected in future GW experiments and can be explored experimentally in laboratories.

\textit{To conclude,} we emphasize, once again, that hunting for apparently unrelated signals from the laboratory and from the sky will help us to break degeneracies beyond the SM theories of cosmology that involve multiple energy scales, and provide the community with a powerful way to search for new physics.

\section{ACKNOWLEDGEMENT}
 The work of SD is supported by the National Natural Science Foundation of China (NNSFC) under grant No. 12150610460. GW acknowledge the STFC Consolidated Grant ST/L000296/1 a.

\bibliography{ref.bib}
\bibliographystyle{unsrt}
\end{document}